\documentclass[aps,prl,floatfix,twocolumn,10pt]{revtex4}

\usepackage{color,amsmath,amssymb,graphicx,latexsym,subfigure}
\usepackage{threeparttable}

\usepackage{titlesec}
\usepackage[toc,title,page]{appendix}
\titleformat{\section}[runin]{\normalfont \bfseries}{\thesection}{1em}{}

\newcommand{\nn}{{\nonumber}}

\newcommand{\beq}{\begin{equation}}
\newcommand{\eeq}{\end{equation}}
\newcommand{\bea}{\begin{eqnarray}}
\newcommand{\eea}{\end{eqnarray}}

\newcommand{\gsim}{\lower.7ex\hbox{$\;\stackrel{\textstyle>}{\sim}\;$}}
\newcommand{\lsim}{\lower.7ex\hbox{$\;\stackrel{\textstyle<}{\sim}\;$}}

\newcommand{\be}{\begin{equation}}
\newcommand{\ee}{\end{equation}}
\newcommand{\ba}{\begin{eqnarray}}
\newcommand{\ea}{\end{eqnarray}}

\newcommand{\pt}{\mathcal{P T}}

\begin{document}

\title{Axion Haloscope Array With $\mathcal{PT}$ Symmetry}

\author{Yifan Chen$^{a}$}
\author{Minyuan Jiang$^{b,a}$}
\author{Yiqiu Ma$^{c}$}
\author{Jing Shu$^{a,d,e,f}$}
\author{Yuting Yang$^{a,d}$}

\affiliation{
$^a$CAS Key Laboratory of Theoretical Physics, Institute of Theoretical
Physics, Chinese Academy of Sciences, Beijing 100190, P.R.China\\
$^b$Weizmann Institute of Science, Rehovot 76100, Israel\\
$^c$Center for Gravitational Experiment, Hubei Key Labolatory for Gravitation and Quantum, School of Physics,
Huazhong University of Science and Technology, Wuhan, 430074, China\\
$^d$School of Physical Sciences, University of Chinese Academy of Sciences, Beijing 100049, China\\
$^e$School of Fundamental Physics and Mathematical Sciences, Hangzhou Institute for Advanced Study, University of Chinese Academy of Sciences, Hangzhou 310024, China\\
$^f$International Center for Theoretical Physics Asia-Pacific, Beijing/Hangzhou, China
}

\begin{abstract}

We generalize the recently proposed $\pt$-symmetric axion haloscope to a larger array with more $\pt$-symmetric structures.
{By broadening the response bandwidth of the signal without increasing the readout noise, the optimized scan rate of the axion haloscope is significantly enhanced, as well as is the signal power.}
 Furthermore, we show that the robustness of the detector towards the variations of the array coupling is the strongest when a \emph{binary tree} structure is introduced which contains a largely enhanced $\pt$ symmetry. The multiple allowed probing sensors can further increase the {scan rate} by a factor of the sensors' number due to the correlation of the signals. This type of array can strongly boost the search for an axion compared to single-mode resonant detection. The enhancement to the {scan rate} becomes the most manifest when applied to the proposed detection using a superconducting radio-frequency cavity with an ac magnetic field where most of the parameter space of the QCD axion above kHz can be probed.

\end{abstract}

\date{\today}

\maketitle

\section{Introduction---}
An axion, with the initial motivation to solve the strong $CP$ problem in QCD~\cite{Peccei:1977hh}, is a strongly motivated hypothetical particle beyond the standard model.
Besides the QCD-axion, axion-like particles (ALPs) also appear generically  in theories with extra dimensions {\cite{Arvanitaki:2009fg}}. 
These particles can be ideal candidates of cold dark matter \cite{Preskill:1982cy, Abbott:1982af, Dine:1982ah}, behaving as a coherent wave within the correlation time and distance.
There are many strategies to search for axion dark matter, for example, using a resonant microwave cavity or superconducting circuit as a haloscope \cite{Sikivie:1983ip, Sikivie:1985yu, Sikivie:2013laa}, an axion-induced birefringence effect for light propagation \cite{Carroll:1989vb,Harari:1992ea,Chen:2019fsq}, and axion-induced nuclear magnetic precession \cite{Graham:2013gfa,Budker:2013hfa,Jiang:2021dby}.
The main experimental platforms include ADMX \cite{Du:2018uak}, SN1987A  \cite{Payez:2014xsa}, and CAST \cite{Anastassopoulos:2017ftl}, which set the current limits to the axion parameters.

The strategy using a resonant microwave system is based on the axion-photon interaction (the so-called inverse Primakoff process) in strong background magnetic fields, which was first discussed by Sikivie\,\cite{Sikivie:1983ip, Sikivie:1985yu}.
The original haloscope includes a microwave cavity embedded in a dc magnetic field, including ADMX\,\cite{Du:2018uak}, ORGAN\,\cite{McAllister:2017lkb}, HAYSTACK\,\cite{Brubaker:2018ebj}, and CAPP\,\cite{Lee:2020cfj}. Generalizations targeted at different axion masses include  superconducting-$LC$ circuits \cite{Sikivie:2013laa} such as dark matter (DM) radio\,\cite{Chaudhuri:2014dla} and  ABRACADABRA\,\cite{Ouellet:2018beu} as well as a superconducting radio-frequency (SRF) cavity embedded in the ac magnetic field \cite{Berlin:2019ahk,Lasenby:2019prg,Lasenby:2019hfz}.

Since the microwave field generated by axion conversion is extremely weak, a microwave resonator with a high-quality factor ($Q$ factor) is essential for the experiment, which at the same time narrows the bandwidth of the detector. Therefore, to search for the axion dark matter in a broad mass spectrum, we need to switch the detector among different central frequencies in the practical running of the detector. The figure of merit for the Sikivie-type axion haloscope with a tunable center frequency is the scan rate $R_a$, defined as\,\cite{Malnou:2018dxn}:
\be
R_a=\int^\infty_0d\Omega\frac{1}{|S_\alpha(\Omega)|^2},\label{Ra}
\ee
where $S_\alpha(\Omega)$ is the sensitivity (noise normalized by the signal amplitude) of the haloscope.
Here, we assume the axion coherence time is short compared to the observation time. {The scan rate of resonant detection is limited by two kinds of noise, i.e., intrinsic fluctuations due to the dissipation of the experimental components and the quantum fluctuations from the readout channel. The resonant detector responds to the signal field in the same way as to the intrinsic fluctuations, 
where the strongest response occurs around the resonant frequency. Thus the frequency domain sensitivity depends on the range that intrinsic fluctuations dominate over the flat readout one, as discussed in detail in  Refs.\;\cite{Chaudhuri:2018rqn,Chaudhuri:2019ntz,Berlin:2019ahk,Lasenby:2019prg,Lasenby:2019hfz}.}

{To improve the scan rate, one can either reduce the system noise level or broaden the detection bandwidth (at the same time not sacrificing the sensitivity), thus responding to a larger off-resonance frequency region for each scan}. Several works targeted at reducing the system noise level\,\cite{Krauss:1985ub, Zheng:2016qjv, Malnou:2018dxn, Chaudhuri:2018rqn, Chaudhuri:2019ntz, Berlin:2019ahk, Lasenby:2019prg, Lasenby:2019hfz, Berlin:2020vrk} have been carried, for example, microwave squeezing technology is useful in improving the scan rate. Recently, a new design\,\cite{Li:2020cwh} which has the feature of $\pt$ symmetry is proposed to substantially broaden the detector bandwidth and thereby significantly reduce the switching time costs. The basic structure of this proposal is a Sikivie-type axion detector assisted with an auxiliary nondegenerate parametric interaction, which was inspired by the white light cavity concept used in laser interferometer gravitational wave detectors\,\cite{Miao:2015pna}.

{In this paper, we proposed different configurations to enhance the {scan rate} further through generalizations of the $\pt$-symmetric design concept to an array of detectors. Since the coherent length of the axion dark matter signal is typically $10^3$ times the Compton length {(see, e.g., Refs.\;\cite{Derevianko:2016vpm,Chaudhuri:2018rqn,Foster:2020fln,Chen:2021bdr})} that a resonant microwave cavity usually matches, one can upgrade it to a fully $\pt$-symmetric setup with multiple probing sensors. On the other hand, variations of experimental parameters from the optimal values can potentially degrade the scan rate enhancement. It turns out that the fully $\pt$-symmetric configuration is more robust against these variations.}

\section{Haloscope with $\pt$ symmetry---}\label{resonatorchain}
\begin{figure}[h]
  	\centering
  	\includegraphics[width=1.0\columnwidth]{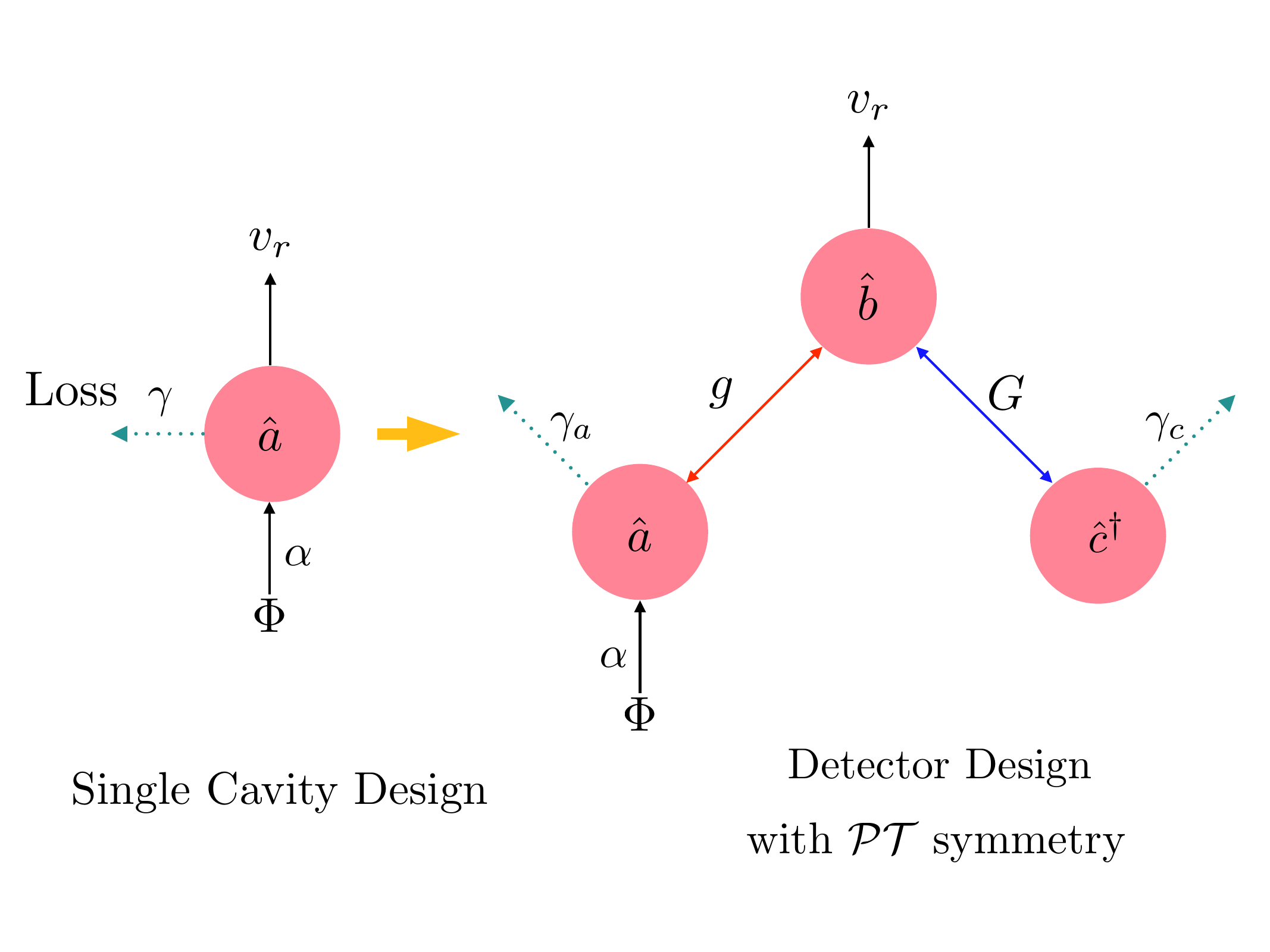}
  	\caption{
  		\footnotesize
  		{Design concept of the detector with $\pt$ symmetry (when $g=G$), where two auxiliary degrees of freedom were introduced with beam-splitter and parametric types of interactions in red and blue, respectively.}}
  	\label{fig:ptscheme}
  \end{figure}

The detector design with a $\pt$-symmetric feature in Ref.\;\cite{Li:2020cwh}, shown in Fig.\;\ref{fig:ptscheme}, can be described by the following  Hamiltonian {in the interaction picture where the free Hamiltonians of the modes are omitted},
\be
\hat{H}_{\textrm{int}} /\hbar= g (\hat a \hat b^\dagger + \hat a^\dagger \hat b) +  G (\hat b \hat c + \hat b^\dagger \hat c^\dagger) +  \alpha (\hat a +\hat  a^\dagger) \Phi,\label{intH}
 \ee
where $\hat a, \hat b, \hat c$ are three cavity modes. The mode $\hat a$ is used to probe the axion signal $\Phi$ with coupling constant $\alpha$, while the mode $\hat{b}$ is connected to the readout port where we extract the detection result. There are two different kinds of interactions in this Hamiltonian, (1) the beam-splitter-type interaction with strength $g$ between modes $\hat a$ and $\hat b$, and (2) the nondegenerate parametric interaction \cite{PCLC, Bergeal} with strength $G$ between modes $\hat b$ and $\hat c$. These modes couple with external continuous fields via
\be
\hat H_{\rm ext}/\hbar=i\sqrt{2\gamma_a}(\hat a\hat u_a^\dag-\hat a^\dag\hat u_a),
\ee
and they take the same form for the $\hat b, \hat c$ modes. 

When $g=G$, the interaction Hamiltonian (\ref{intH}) reduces to:
\be
\mathcal{H}_{\textrm{WLC}}/\hbar =  g (\hat a+\hat c^\dag)\hat b^\dag+ g (\hat a^\dag+\hat c)\hat b+ \alpha (\hat{a} + \hat{a}^\dagger) \Phi ,
\ee
and the $\pt$ symmetry emerges. In this case, $\mathcal{P}$ transformation switches $\hat a$ and $\hat c$ and $\mathcal{T}$ transformation switches the creation and annihilation operators. Under the joint $\pt$ transformation, the Hamiltonian keeps its original form. Moreover, in the ideal case when $\hat a/\hat c$ are lossless, the response of the $\hat a+\hat c^\dag$ to the axion driving is inversely proportional to the frequency, thereby achieving a significant enhancement of the  signal, thus sensitivity $S_\alpha(\Omega)$ at low frequencies. The conceptual designs and how the system behaves under the non-ideal situation where the $\hat a/\hat c$ possess internal loss and $g\neq G$ has been demonstrated and optimized in Ref.\;\cite{Li:2020cwh}.  

{In the following sections, we show that the sensitivity can be further enhanced in systems with enlarged $\pt$-symmetric structures.}


\section{A chain of $\pt$-symmetric detector---} 
\begin{figure}[h]
  	\centering
  	\includegraphics[width=1.1\columnwidth]{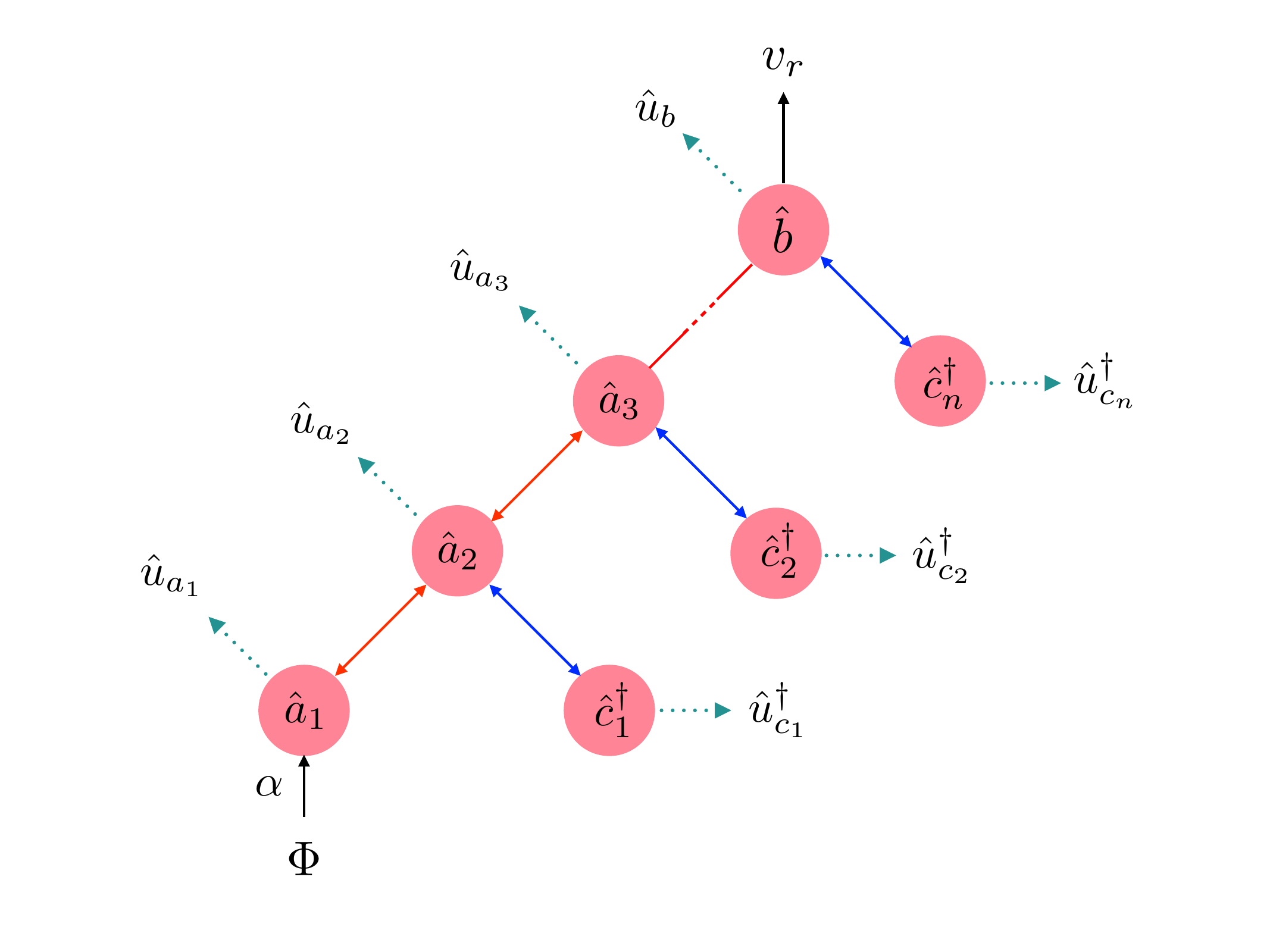}
  	\caption{
  		\footnotesize
  		{Chain detector configuration, where the red lines represent the beam-splitter type of interaction while the blue lines represent the parametric type of interaction. Each circle represents a degree of freedom of the system and only $\hat a_1$ is driven by the axion signal field. The information of the axion field is detected at the $\hat b$ port.}}
  	\label{fig:chaindetector}
  \end{figure}

The advantage of the $\pt$-symmetric detector design comes from its high response to the signal at low frequencies. Such an advantage can be further enhanced by considering an extension of the model with more $\pt$-symmetric structures which we named the \emph{chain detector design}, as shown in Fig.\;\ref{fig:chaindetector}. The interaction Hamiltonian of this chain detector design is
\ba
 \hat H_{\textrm{CD}} /\hbar=&\sum^n_{i=1} g\hat a_i\hat a^\dag_{i+1}+G\hat c_i\hat a_{i+1}\nn
 \\&+g\hat a_n\hat b^\dag+G\hat c_n\hat b+\alpha\hat a_1\Phi+{\rm h.c}.
 \ea
In the case of $g=G$, we have: $g\hat a_i\hat a^\dag_{i+1}+G\hat c_i\hat a_{i+1}\rightarrow g(\hat a_i+\hat c^\dag_i)\hat a_{i+1}$ and define the $\pt$-invariant mode $\hat A_i \equiv \hat a_i+\hat c_i^\dag$.
In the lossless case when the external continuous electromagnetic bath $\hat u_r$ only couples to the readout $\hat b$ mode via strength $\sqrt{2\gamma_r}$, we have the following equations of motion:
\ba
&\dot{\hat A}_i=-ig\hat A_{i-1},\quad\dot{\hat A}_1=-i\alpha\Phi,\nn\\
&\dot{\hat b}= -\gamma_r \hat b - i g \hat A_n+ \sqrt{2\gamma_r} \hat u_r,
&\hat v_r= \hat u_r - \sqrt{2\gamma_r}\hat b.
\ea
In the frequency domain, these equations can be  stacked in the following way,
\be\hat v_{r}(\Omega)=\frac{\Omega-i\gamma_r}{\Omega+i\gamma_r}\hat u_r(\Omega) - \frac{i \sqrt{2\gamma_r} \alpha \Phi}{\Omega+i\gamma_r}\left(\frac{g}{\Omega}\right)^{n},\ee
{where $\Omega$ is the frequency shift from the resonant frequency $\omega_{\rm rf}$.}
Clearly, there is an enhanced amplification factor $\mathcal{G}^{n}(\Omega)=(g/\Omega)^{n}$ due to the chain structure. Thus the low-frequency signal response scales as $\sim\Omega^{-n}$ \footnote{Note that although the signal response gets amplified at low frequencies, the signal decreases faster at high frequency region.}, while at the same time the noise is still at the shot-noise level since $S_{u_ru_r}=1$ (in principle, this shot-noise level can be further reduced via squeezing technology so that $S_{u_ru_r}=e^{-2r_s}$, where $r_s$ is the squeezing degree). The signal power and noise spectrum in this ideal lossless case are given by:
\ba\frac{1}{S_\alpha(\Omega)}=\frac{2\gamma_{r}\alpha^{2} S_{\Phi} (\Omega)}{\gamma_{r}^{2}+\Omega^{2}}\left(\frac{g^{2}}{\Omega^{2}}\right)^{n}e^{2r_s}.\ea

\section{Imperfections and robustness---} In practice the lossless condition and the $\pt$-symmetric condition can not be achieved as ideal, and the robustness of the device's sensitivity  towards varying the loss rate and $\pt$-symmetric condition needs to be taken into account. Let us first consider the case when $g=G$, but there exists an internal loss $\gamma_{a/c}$ and we assume that the internal loss for each resonator is the same so that $\gamma_a=\gamma_c=\gamma$. The signal enhancement factor in this case would be $\mathcal{G}(\Omega)=g^2/(\gamma^2+\Omega^2)$ and the signal power $S^{\textrm{CD}}_{\textrm{sig}} (\Omega) $, so the noise spectrum  due to the readout $S^{\textrm{CD}}_{\textrm{r}} (\Omega) $ and the internal loss $S^{\textrm{CD}}_{\textrm{int}}$ now can be written respectively as:
\begin{align}
S^{\textrm{CD}}_{\textrm{sig}} (\Omega) &= \frac{2\gamma_{r}\alpha^{2} S_{\Phi} (\Omega)}{(\gamma+\gamma_{r})^{2}+\Omega^{2}}\mathcal{G}^{n}(\Omega)\label{sigp},\\
S^{\textrm{CD}}_{\textrm{r}} (\Omega) &= \frac{(\gamma-\gamma_{r})^{2}+\Omega^{2}}{(\gamma+\gamma_{r})^{2}+\Omega^{2}}S_{u_r}\label{Sr},\\
S^{\textrm{CD}}_{\textrm{int}}(\Omega) &= \frac{4\gamma\gamma_r}{(\gamma+\gamma_r)^2+\Omega^2}\left[S_{u_b}+\sum^n_{i=1}\mathcal{G}^{n+1-i}(S_{ua_i}+S_{uc_i})\right],
\label{Sint}
\end{align}
where the noise spectra $S_{ua_i},S_{uc_i}$ are the loss contributions from the $\hat a_i$ and $\hat c_i$, including both vacuum and thermal fluctuations. Even at $\Omega < \gamma$, as long as $g\gg\gamma$, there is still a significant amplification of the signal power.  We can vary the value of $\gamma_r$ while fixing $g=G$ to optimize the scan rate in Eq. (\ref{Ra}) and the optimized spectrum for $n=2,3$ is shown in the Appendix. It turns out that increasing the resonant chain levels could improve the scan rate $R_a$ by a factor of $\left(g/\gamma n_{\textrm{occ}}\right)^{2n/(2n+1)}$, through broadening the range of $S^{\textrm{CD}}_{\textrm{int}}(\Omega) \gg S^{\textrm{CD}}_{\textrm{r}}(\Omega)$, where $n_{\textrm{occ}}$ is the thermal occupation number.

\begin{figure}[h]
  	\centering
  	\includegraphics[width=1.0\columnwidth]{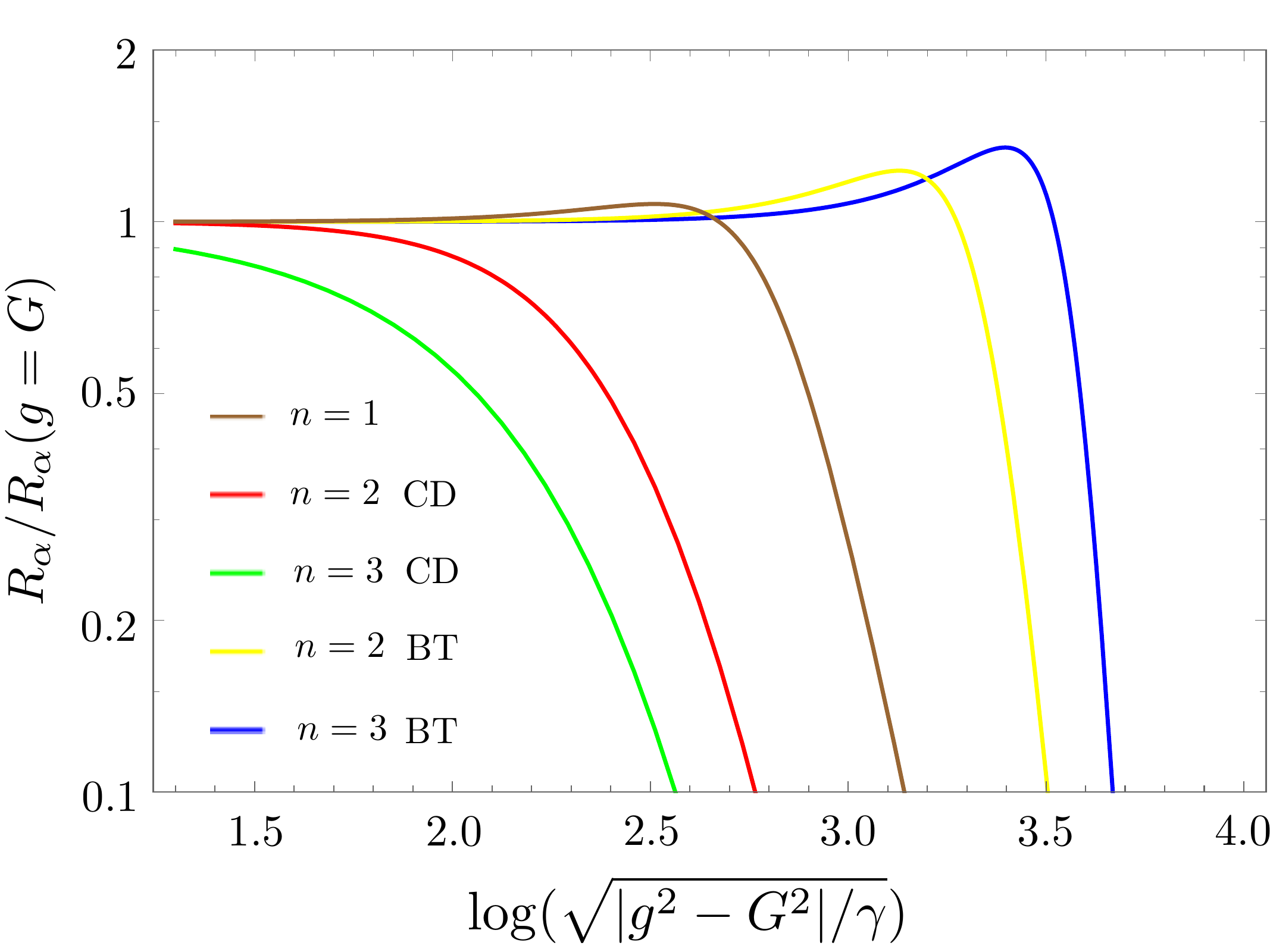}
  	\caption{
  		\footnotesize
  		{The effect of $\pt$ symmetry breaking when $g^2-G^2 > 0$ to the scan rate of the \emph{chain detector} configuration and \emph{binary tree} configuration at zero temperature for $g/\gamma = 10^4$.}}
  	\label{fig:g-robustness}
  \end{figure}

On the other hand, the mismatch between $g$ and $G$, which breaks the $\pt$ symmetry, will affect the system response as well and it has been also discussed in Ref.\;\cite{Li:2020cwh}.  Therefore we also need to test the robustness of the optimized scan rate, and the result is shown in Fig.\;\ref{fig:g-robustness}, where we calculate how the scan rate ratio $R_\alpha(g\neq G)/R_\alpha(g=G)$ would change with respect to $\sqrt{g^2-G^2}$ at zero temperature.  In this figure, the scan rate would drop by half if $g^2-G^2\sim 10^5\gamma^2$ when $g/\gamma = 10^4$, which requires $g - G \ll 5 \gamma$. For the system with large $g$, it could be a fine-tuning problem and thereby the system is less robust towards the $g-G$ mismatch.

\section{Binary tree design---}
\begin{figure}[h]
  	\centering
  	\includegraphics[width=1.0\columnwidth]{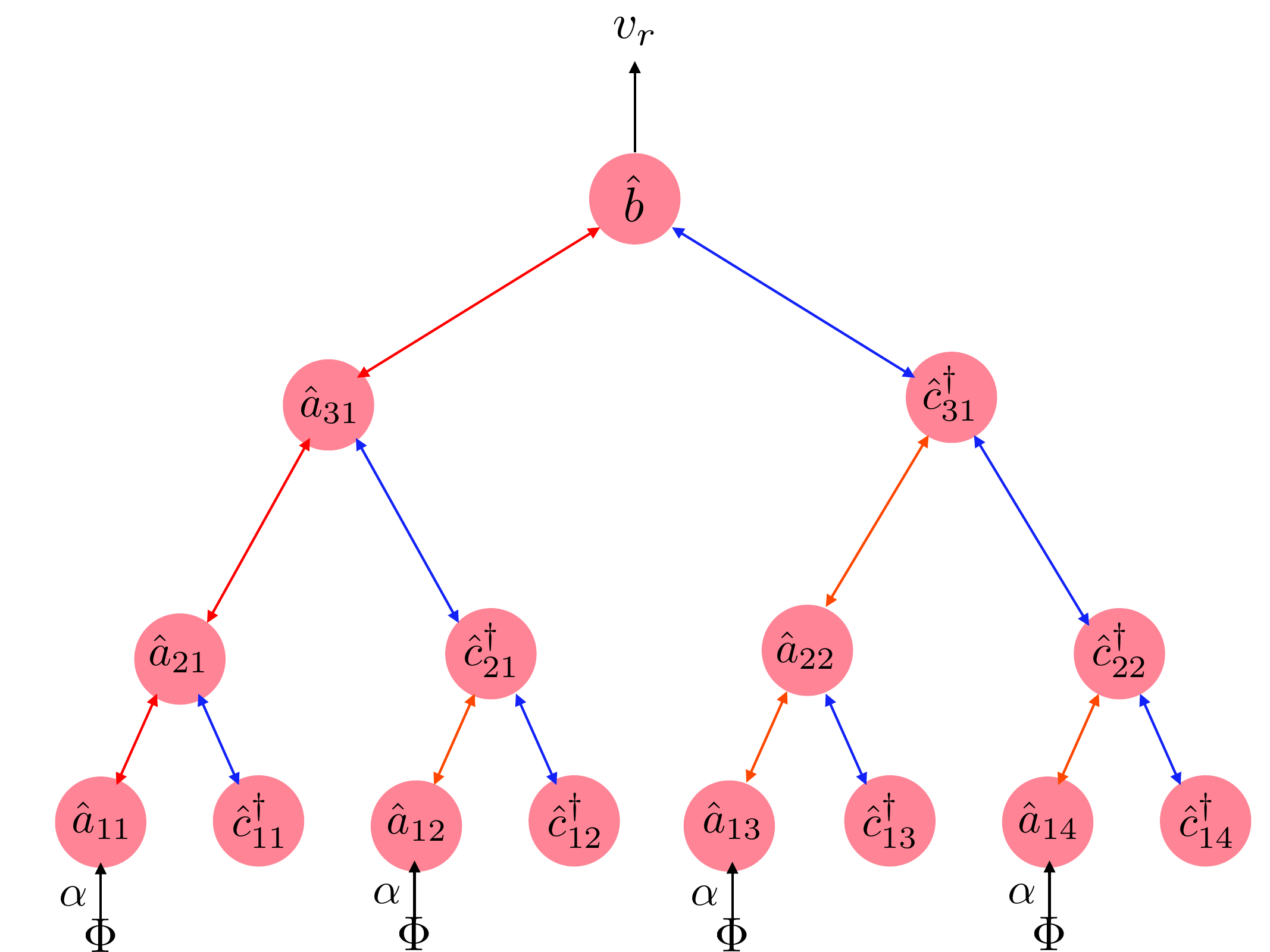}
  	\caption{
  		\footnotesize
  		{An example of a  \emph{binary tree design} with $n = 3$, with red lines denoting a beam-splitter-type interaction and blue lines denoting a parametric-type interaction. The probing modes to the dark matter are chosen to be the $\hat a_{1i}$ modes in the lowest level.}}
  	\label{FBT}
  \end{figure}

An alternative design, with the schematic diagram shown in Fig.\;\ref{FBT} (in this paper, we call it the \emph{binary tree design}), could relieve the problem of robustness shown in the previous section. The corresponding Hamiltonian can be written as:
\be
\begin{split}
\hat{H}_{\textrm{BT}}/\hbar&=(g\hat a^\dag_{n1}+G\hat c_{n1})\hat b+\sum^{2^{n-1}}_{j=1}\alpha\hat a_{1j}\Phi\\
&+\sum^n_{i=2}\sum^{2^{n + 1 - i}}_{j=1} \hat a_{ij}( g \hat a^\dag_{i-1,2j-1}+ G\hat c_{i-1,2j-1})\\
&+\sum^n_{i=2}\sum^{2^{n + 1 - i}}_{j=1} \hat c_{ij}( g \hat a^\dag_{i-1,2j-1}+ G\hat c_{i-1,2j}) + h.c.\label{HBT}
\end{split}
\ee
It is important to note that in this design, we are allowed to have multiple resonant modes to probe the axions, while in the chain detector, we only have one. Since the typical correlation wavelength of the axion dark matter is much larger than the spatial scale between theses resonators, there is a coherent enhancement of signal fields as well as the scan rate. However, we need a benchmark to compare different device designs \footnote{Another example is in the gravitational wave detectors, to show the advantage of detector upgrades, one usually needs to compare the sensitivity curve of the upgraded detector and the original detector assuming the same intra-cavity power.}, and such a benchmark is chosen to be the total signal amplitude at the input. This is to say that when we compare the \emph{chain resonator design} and the \emph{binary tree design}, we set the total signal amplitude to be the same. Practically speaking, it means we can reduce the strength of the background magnetic field in the coupling cavities $\hat a_{1i} (i=1,...,N)$ by a factor of $N$. In this sense, there is no advantage of the \emph{binary tree design} over the chain detector design when there is no loss and $\pt$ symmetry is strict.

The structure of the interaction between $\hat a_{ij}$ and $\hat c_{ij}$ renders the Hamiltonian (\ref{HBT}) to have a largely enhanced $\pt$ symmetry. It turns out to demonstrate much higher robustness to the $g - G$ mismatch  than the resonant chain detector as we can see from Fig.\;\ref{fig:g-robustness}, and increasing the level $n$ of the \emph{binary tree design} also enhances the robustness.

\begin{figure}[h]
  	\centering
  		\includegraphics[width=1.0\columnwidth]{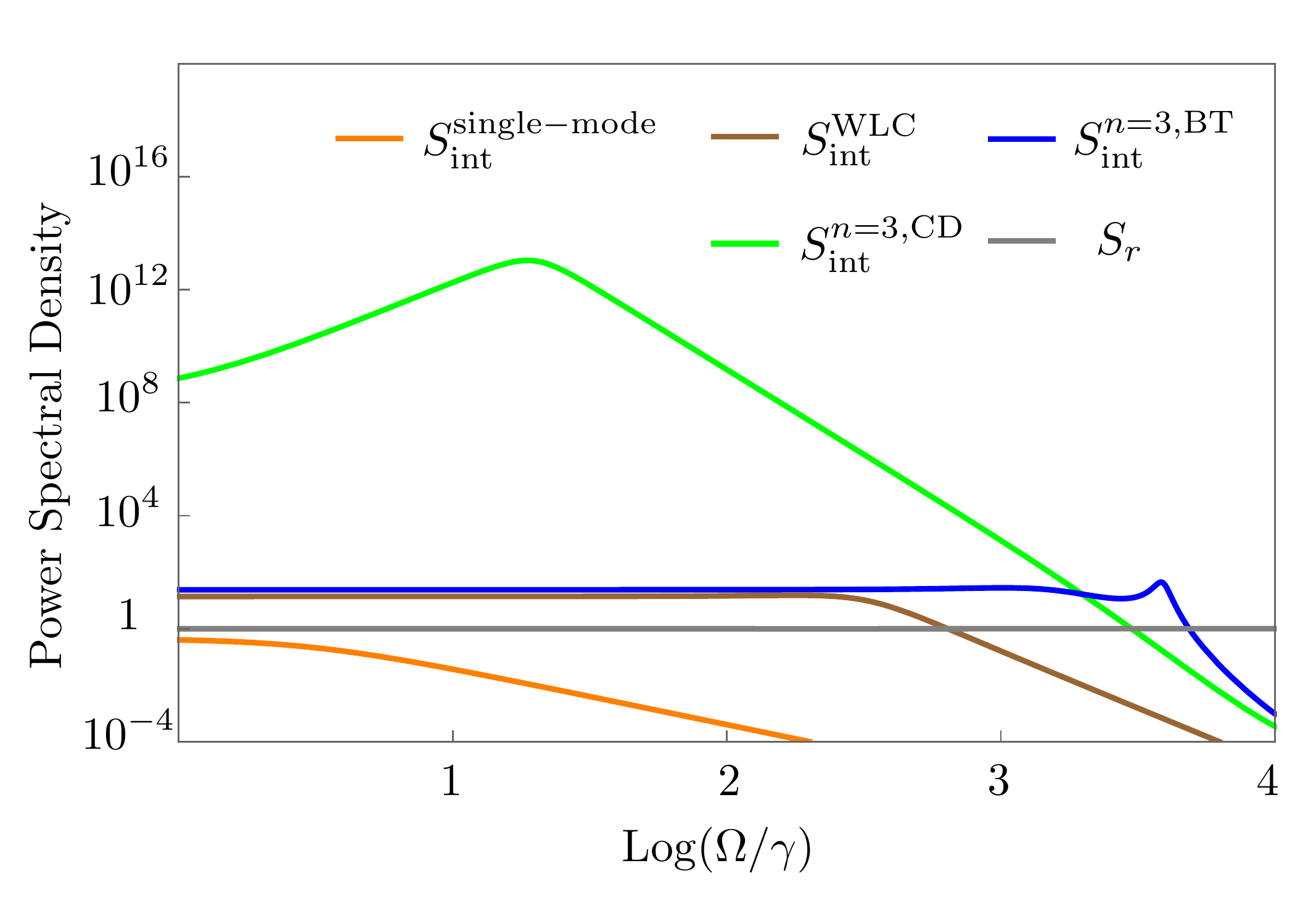}
  	\caption{
  		\footnotesize
  		{{The corresponding noise PSD for $n = 1$ and $3$ of a \emph{chain detector} and \emph{binary tree} at $g/\gamma = 10^4$ when both $\gamma_r$ and $g^2 - G^2$ are tuned to give the optimized scan rate. 
  		The PSDs of a binary tree are flatter compared to the one of the chain detector. 
  		In comparison, we also show the PSD for the single-mode resonator. Each scan can probe axion dark matter with mass $\omega_{\rm rf} \pm \Omega$.}}}
  	\label{nPToptPSD}
  \end{figure}

{In the Appendix, we numerically optimize the scan rate in terms of both $\gamma_r$ and $g^2 - G^2$ , and the corresponding noise power spectral density (PSD) is in Fig.\;\ref{nPToptPSD}. The range where internal loss noise $S_{\textrm{int}}$ dominates over the readout noise $S_{\textrm{r}}$ is indeed broadened once the level $n$ increases, contributing to the enhancement of the scan rate compared with the case of the single-mode resonator. 
Take the array extension of a typical Sikivie design concept  (where a microwave cavity is embedded in a dc magnetic field for axion detection) as an example:
Each scan can effectively probe the axion mass within  $\omega_{\rm rf} \pm \Delta\omega_{\textrm{sc}}$, where {$\Delta\omega_{\textrm{sc}} \simeq (g^{2n} \gamma n_{\textrm{occ}})^{1/(2n+1)}$ is the effective scan bandwidth in which $S_{\textrm{int}}(\Omega) \gg S_{\textrm{r}}(\Omega)$ }}. {Notice that the flat distribution of the noise PSD for the \emph{binary tree} is another feature of robustness compared with the \emph{chain detector}}.

\section{Experimental expectations---}
We now discuss the application of the $\pt$-symmetric array to the electromagnetic resonant detectors. These include a microwave cavity embedded in a strong dc magnetic field with the coupling strength $\alpha=g_{\Phi\gamma} \eta B_0 \sqrt{m_\Phi V}$, where $g_{\Phi\gamma}$ is the axion-electromagnetic-field coupling, $\eta$ is the overlapping factor of the cavity mode with the background magnetic field $B_0$, and $V$ is the cavity volume \,\cite{Sikivie:1983ip, Sikivie:1985yu}. We also consider superconducting-$LC$ circuits \cite{Sikivie:2013laa}/superconducting radiofrequency (SRF) cavity \cite{Berlin:2019ahk, Lasenby:2019prg, Lasenby:2019hfz} embedded in the dc/ac magnetic field with corresponding coupling strengths to be $\alpha_{\rm LC} = g_{\Phi\gamma} B_0 V^{5/6} m_\Phi^{3/2}$ and  $\alpha_{\textrm{RF}} = g_{\Phi\gamma} \eta B_0 m_\Phi \sqrt{V/\omega_{\rm rf}}$ respectively, where $\omega_{\rm rf}$ is the resonant frequency of the SRF cavity.

\begin{figure}[h]
    \centering
    \includegraphics[width=1.0\columnwidth]{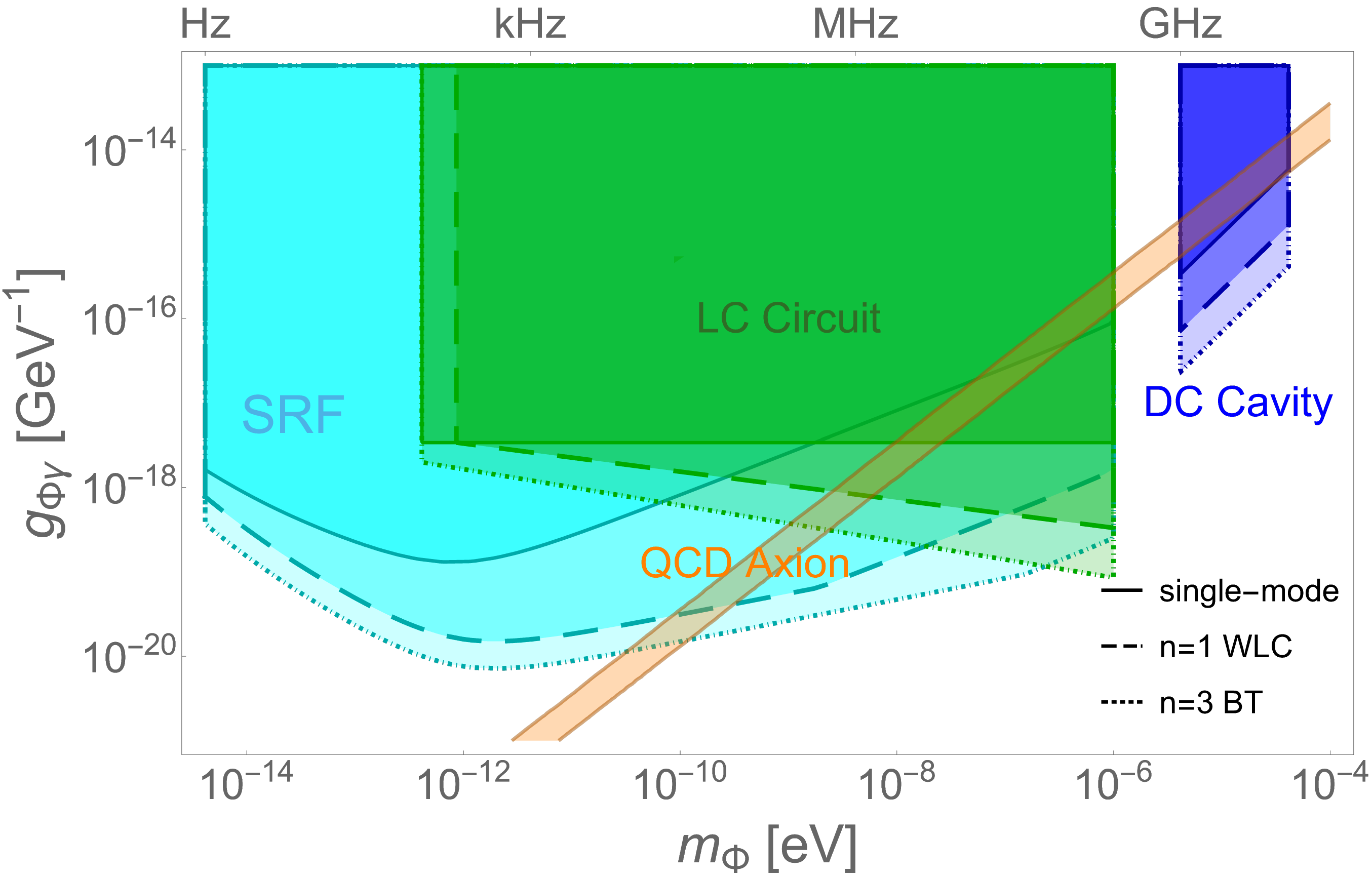}
    \caption{Physics reach for cavity, $LC$ circuit with dc magnetic field, and SRF with ac magnetic field. Here, we choose the \emph{binary tree} detector design as an example (because of its robustness to the $g - G$ mismatch).  {The scaling with $m_\Phi$ changes at low frequency for SRF due to the phase fluctuation noise \cite{Berlin:2019ahk} and the scan bandwidth saturating $m_\Phi$.}}
    \label{PR}
\end{figure}

{We consider the \emph{binary tree design} only due to its robustness, as discussed in the previous sections.}
The potential physics reaches based on the three different types of experiments are shown in Fig.\;\ref{PR} with the integration time distributed for  each $e$-fold in the axion mass to be $t_e = 10^7$ s and the signal-to-noise ratio (SNR) reaching $1$. 
In the Appendix, we show that the SNR is proportional to the square root of the scan rate in Eq.\;(\ref{Ra}). Since typically $g$ cannot be larger than the resonant frequency $\omega$, we take $g=Q_{\textrm{int}}\gamma$. Thus the high-quality factor as well as an almost fixed thermal occupation number of the SRF cavity leads to a much larger enhancement. The benchmark parameters for a cavity with a dc magnetic field is $V = 1\ \textrm{m}^3$, $\eta = 1$, $B_0 = 4$ T, $T = 10 $ mK, $Q_{\textrm{int}} = 10^4$ while the $LC$ circuit only differs by $Q_{\textrm{int}} = 10^6$ \cite{Chaudhuri:2018rqn}. For SRF, the differences with the traditional cavity are $Q_{\textrm{int}} = 10^{12}$, $B_0 = 0.2$ T, $T = 1.8$ K, and the resonant frequency is almost fixed to be $\omega_{\textrm{rf}} = 2\pi\;\textrm{GHz} + m_{\Phi}$ \cite{Berlin:2019ahk}. Below kHz, we also include the contribution of the phase fluctuation noise that dominates over the thermal noise, with the overlapping factor between the pumping magnetic field and the signal electric field to be $\epsilon_{\textrm{1}d} = 10^{-5}$ and the quality factor of the pumping field to be the same as $Q_{\textrm{int}}$. {We require the scan bandwidth $\Delta\omega_{\rm sc}$ to be no larger than the axion mass $m_\Phi$, as discussed in the Appendix, which makes the scaling of the physics reach change below $10$ MHz.}

{The non-degenerate interactions for these three systems are already experimentally realized in Refs.\;\cite{PCLC, Bergeal}. There are still several potential challenges to consider for the practical implementation. 
The first one comes from the compatibility between the strong magnetic field and the non-degenerate interaction. Since the realizations of such a type of interaction usually require superconducting ingredients, a spatial separation away from the magnetic background is necessary to maintain the superconductivity. A conducting wire from the cavity/circuit can solve this issue, as mentioned by Ref.\;\cite{Wurtz:2021cnm}. 
Another potential obstacle comes from the phase fluctuation of the pumping mode to realize the non-degenerate interaction, leading to a time-dependent value of the coupling $G$ \cite{WBAR}. The binary tree is more robust against such variations as well.}
Finally, the transmission losses due to the interactions between different resonant modes will modify the intrinsic dissipation coefficients $\gamma$, thus requiring more precise calibration for the modes at the lowest level.

\section{Conclusions---} In this work, we explore the generalization of the axion detector design assisted with a $\pt$-symmetric quantum amplifier to be a multiple resonant system for the further enhancement of sensitivity. {With comparable parametric  coupling and beam-splitter coupling, a $\pt$-invariant mode $\hat{a} + \hat{c}^\dagger$ is formed, which can transform the signal to the next connected mode without suppression of the signal response. Finally, the signal response is further enhanced by flowing through different $\pt$-invariant pairs while the readout noise response function stays the same.}
Two generalized scheme configurations are discussed and compared  respectively: the \emph{chain detector} configuration and the \emph{binary tree} configuration in terms of the signal and noise response as well as the robustness towards the variation of the optimized system parameters.

We found that both of these detector configurations have the potential capability of a scan rate enhancement {by broadening the bandwidth of the signal response without increasing the readout noise. 
Considering the variation of experimental parameters, we show that the \emph{chain detector} configuration is less robust toward  $\pt$ symmetry breaking than the \emph{binary tree} configuration.}

{For an electromagnetic resonant system such as a cavity or $LC$ circuit, these $\pt$-symmetric couplings have been already achieved in Refs.\;\cite{PCLC, Bergeal}, thus can be directly applied to most current experiments/proposals.} These improvements of detector capability for constraining the axion mass and the axion-photon coupling of the design concept in this work for three different types of electromagnetic resonant experiments (cavity, $LC$ circuit with a static magnetic field, and SRF with an ac magnetic field) are also discussed. {The enhancement to the scan rate can approach $\sim Q_{\textrm{int}}/ n_{\textrm{occ}}$ for static
field experiments, or for SRF searches at sufficiently high axion masses.
The high-quality factor $Q_{\textrm{int}}$ of the SRF thus can lead to a significant enhancement and probe most of the QCD axion parameter space above kHz. }

\section*{Acknowledgements} 

We are grateful to Yanbei Chen, Saptarshi Chaudhuri, Konrad W. Lehnert and Yue Zhao for useful discussions. This work is supported by the National Key Research and Development Program of China under Grant No. 2020YFC2201501. 
Y.C. is supported by the China Postdoctoral Science Foundation under Grants No. 2020T130661 and No. 2020M680688, the International Postdoctoral Exchange Fellowship Program, and by the National Natural Science Foundation of China (NSFC) under Grant No. 12047557.
The research of M.J. is supported by a research grant from Dr. Adrian Langleben, the Veronika A. Rabl Physics Discretionary Fund, and the Estate of Rachel Berson. Y.M. is supported by the university start-up funding provided by Huazhong University of Science and Technology. J.S. is supported by the National Natural Science Foundation of China under Grants No. 12025507, No. 12150015, and No. 12047503, and is supported by the Strategic Priority Research Program and Key Research Program of Frontier Science of the Chinese Academy of Sciences under Grants No. XDB21010200, No. XDB23010000, and No. ZDBS-LY-7003, and CAS project for Young Scientists in Basic Research YSBR-006. Y.C. would like to thank the Center for Gravitational Experiment for their kind hospitality.

\appendix
\renewcommand{\theequation}{A-\arabic{equation}}
  \setcounter{equation}{0}
\section*{Appendix: Optimization of signal-to-noise ratio}\label{optimization}
In Refs.\;\cite{Chaudhuri:2018rqn,Chaudhuri:2019ntz,Berlin:2019ahk,Lasenby:2019prg,Lasenby:2019hfz}, it was shown that the total signal-to-noise ratio (SNR) for a square-law detection, with an optimized filter chosen, is proportional to 
\ba \textrm{SNR}^2 (\gamma_r) &=& t_{\textrm{int}} \int_{- \infty}^{+ \infty} d\Omega \left(\frac{S_{\textrm{sig}}}{S_{\textrm{N}}} (\Omega, \gamma_r)\right)^2\nn\\
&=& t_e \frac{\Delta \omega_{\rm sc}}{m_\Phi} \Delta \omega_s \left(\frac{S_{\textrm{sig}}}{S_{\textrm{N}}} (0, \gamma_r)\right)^2\nn\\
&=& \frac{t_e \rho^2_{\textrm{DM}} Q_\Phi}{m_\Phi^6} \int_{- \infty}^{+ \infty} d\Omega \left(\frac{S_{\textrm{sig}}}{S_{\textrm{N}} S_\Phi} (\Omega, \gamma_r)\right)^2,\nn\\ \label{SNRS}\ea
where  $S_{\textrm{N}} \equiv S_{\textrm{int}} + S_{\textrm{r}}$ is the total noise power spectral density (PSD). It is assumed that the scan is performed uniformly in $\log m_\Phi$, with the same time $t_e$ distributed for each $e$-fold in axion mass. $t_{\textrm{int}} = t_e \Delta \omega_{\rm sc}/m_\Phi$ is the integration time within one scan step, required to be larger than the cavity ring-up time-scale and dark matter coherent time $2\pi Q_\Phi/m_\Phi$. $\Delta \omega_{\rm sc}$ is defined as the single scan step within which the expected SNR is an $O (1)$ of the maximum value and $\Delta \omega_s$ is the bandwidth of $\left(S_{\textrm{sig}} / S_{\textrm{N}} \right)^2$. In the case where the dark matter bandwidth $m_\Phi/Q_\Phi$ is smaller than the sensitivity width of the detector, i.e., the width of $\left[ S_\textrm{sig} /(S_{\textrm{N}} S_\Phi) \right]^2$, one has $\Delta \omega_s = m_\Phi/Q_\Phi$ and $\Delta \omega_{\rm sc}$ is the sensitivity width. Now it is clear that the quantity needed to be optimized is the integral in the last line of Eq.\;(\ref{SNRS}), which is the scan rate in Eq.\;(\ref{Ra}). We parametrize the SNR to be 
\be \textrm{SNR} = \frac{\rho_{\textrm{DM}} \alpha^2}{m_\Phi^{3}} \sqrt{\frac{Q_\Phi t_e}{\gamma}}\ \widetilde{\textrm{SNR}}.\label{SNRpara}\ee
 
\begin{figure}[h]
  	\centering
  	\includegraphics[width=0.8\columnwidth]{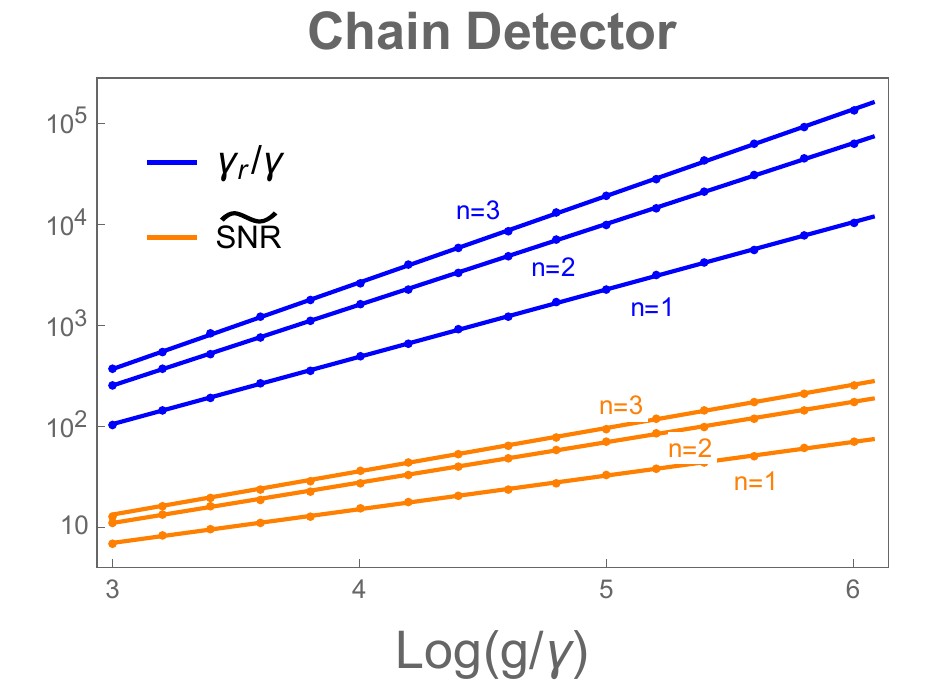}\\
  	  	\includegraphics[width=0.8\columnwidth]{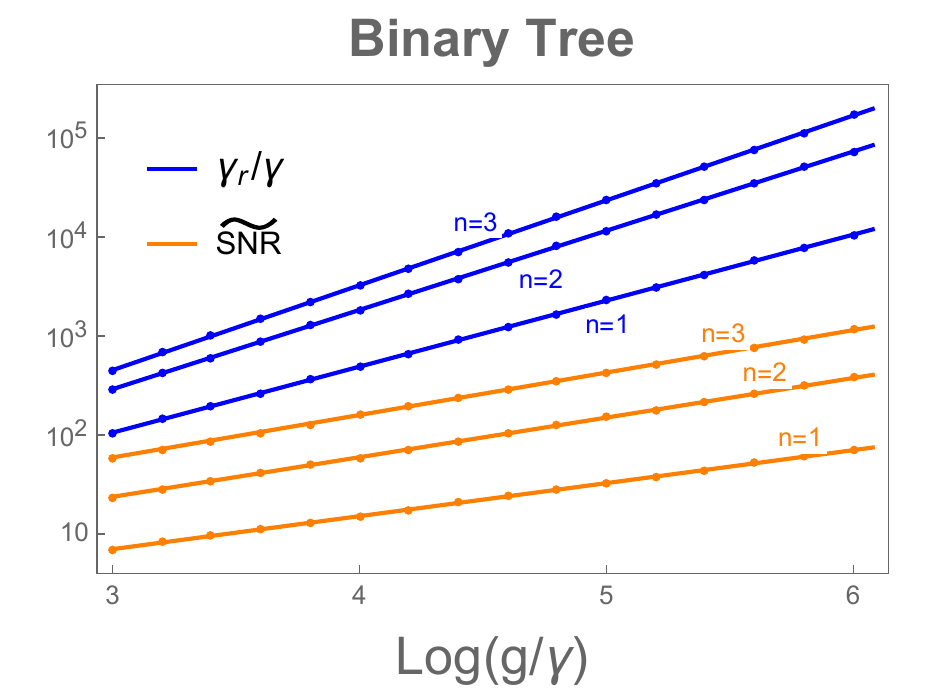}\\
  	  	
  	\caption{
  		\footnotesize
  		{Optimization conditions for $\gamma_r$ when $g=G$ and corresponding $\widetilde{\textrm{SNR}}$ for chain detector and binary tree haloscope, shown also in Eqs.\;(\ref{NoptRC}) and (\ref{noptBT23}).}}
  	\label{RCoptS}
  \end{figure}

 For the chain detector case, using Eqs.\;(\ref{sigp})--(\ref{Sint}), we
 numerically maximize $\widetilde{\textrm{SNR}}$ with $n = 1, 2, 3$ at zero temperature and fit the optimization conditions for $\gamma_r$ and the corresponding $\widetilde{\textrm{SNR}}$:
\ba
n=1&:& \gamma_r=1.05 (g^2\gamma)^{1/3}, \widetilde{\textrm{SNR}}=0.70 (g/\gamma)^{1/3},\nn\\
n=2&:& \gamma_r=1.01 (g^4\gamma)^{1/5}, \widetilde{\textrm{SNR}}=0.70 (g/\gamma)^{2/5},\nn\\
n=3&:& \gamma_r=0.99 (g^6\gamma)^{1/7}, \widetilde{\textrm{SNR}}=0.70 (g/\gamma)^{3/7}.\nn\\
\label{NoptRC}\ea
The results are shown in the upper panel of Fig.\;\ref{RCoptS}.

  \begin{figure}[h]
  	\centering
  	\includegraphics[width=1.0\columnwidth]{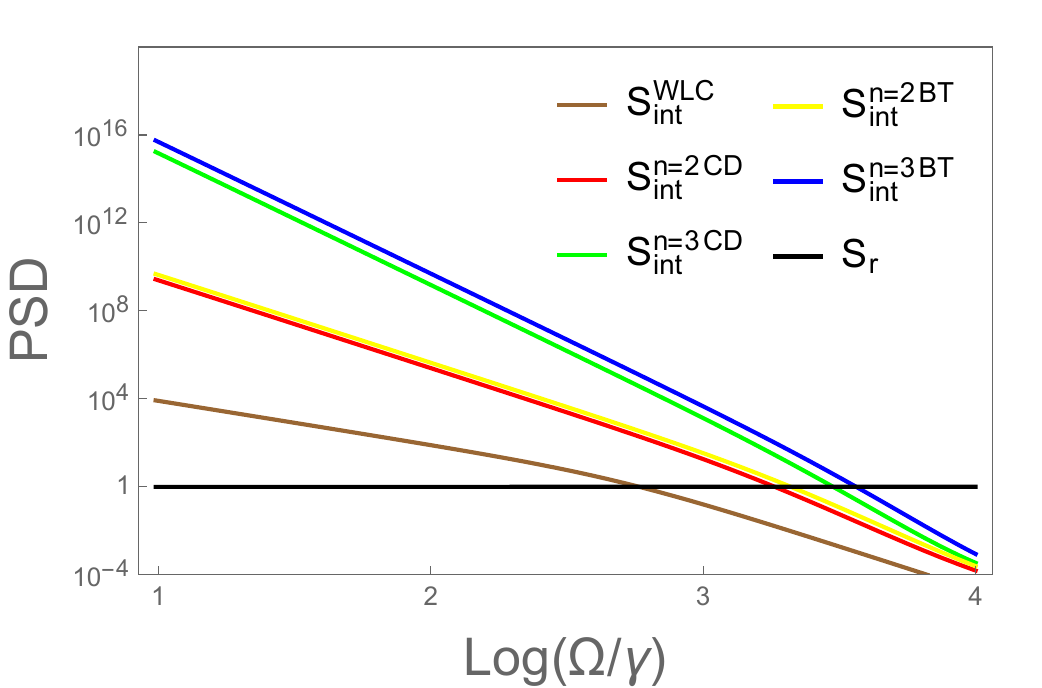}
  	  	
  	\caption{
  		\footnotesize
  		{The corresponding noise PSD for $n = 1, 2, 3$ at $g/\gamma = 10^4$ under the optimization condition in Fig.\;\ref{RCoptS}. $S_{\textrm{r}}$ for the three are almost the same.}}
  	\label{PToptPSD}
  \end{figure}
  
  To understand and generalize the optimized $\widetilde{\textrm{SNR}}$ analytically, one first takes $\gamma_r \simeq (g^{2n} \gamma)^{1/(2n+1)}$ according to the optimization conditions in Eq.\;(\ref{NoptRC}). This makes $S_{\textrm{r}}$ in Eq.\;(\ref{Sr}) to be a flat spectrum, also shown in Fig.\;\ref{PToptPSD}. If the relevant bandwidth in Eq.\;(\ref{SNRS}) is smaller  than $g$, $S_{\textrm{int}}$ is dominated by the contribution from the lowest level, i.e., $S_{u_{a_1}}$ and $S_{u_{c_1}}$ terms in Eq.\;(\ref{Sint}), and
  \be S_{\textrm{int}}(\Omega \ll g) \simeq \frac{4\gamma\gamma_{r}}
{(\gamma+\gamma_{r})^{2}+\Omega^{2}}\left(\frac{g^2}{\gamma^2+\Omega^2}\right)^{n}.\ee  We can now see that the sensitivity bandwidth $\Delta \omega_{\rm sc}$ in Eq.\;(\ref{SNRS}) is just the range that intrinsic fluctuations $S_{\textrm{int}}$ dominates over the readout noise $S_{\textrm{r}}$,
\be S_{\textrm{int}}(\Omega < \Delta\omega_{\rm sc}) > S_{\textrm{r}}(\Omega < \Delta\omega_{\rm sc})\simeq 1\label{SintgSr},\ee
since $\left[S_{\textrm{sig}} / (S_{\textrm{int}} S_\Phi) \right]^2$ remains a constant within the bandwidth. The inequality in Eq.\;(\ref{SintgSr}) leads to
\be \Delta\omega_{\rm sc} \simeq \gamma_r \simeq (g^{2n} \gamma)^{\frac{1}{2n+1}} < g.\ee
Taking it back to Eq.\;(\ref{SNRS}), one has
\be \textrm{SNR} \simeq  \left(\frac{g}{\gamma}\right)^{\frac{n}{2n+1}} \frac{\rho_{\textrm{DM}} \alpha^2}{m_\Phi^{3}} \sqrt{\frac{Q_\Phi t_e}{\gamma}},\ee 
which matches well with our numerical optimization in Eq.\;(\ref{NoptRC}). For large $n$, this leads to an enhanced SNR by a factor approaching $\left(g/\gamma\right)^{1/2}$.

For the binary tree case, taking $g = G$ and all the intrinsic dissipation coefficients to be universal, one can calculate the signal and the noise PSD,
\ba S^{\textrm{BT}}_{\textrm{sig}} (\Omega) &=& \frac{\gamma_{r}\alpha^{2} S_{\Phi} (\Omega)}{2[(\gamma+\gamma_{r})^{2}+\Omega^{2}]}\left(\frac{4g^{2}}{\gamma^{2}+\Omega^{2}}\right)^{n},\\
S^{\textrm{BT}}_{\textrm{int}}(\Omega) &=& \frac{4\gamma\gamma_{r}}{(\gamma+\gamma_{r})^{2}+\Omega^{2}}\Big[ S_{u_{b}}\nn\\
&+&\sum_{i=1}^{n} \sum_{j=1}^{2^{n+1-i}} \left(\frac{g^{2}}{\gamma^{2}+\Omega^{2}}\right)^{n + 1 -i} S_{u_{a_{ij}}}\nn\\
&+&\sum_{i=1}^{n} \sum_{j=1}^{2^{n+1-i}} \left(\frac{g^{2}}{\gamma^{2}+\Omega^{2}}\right)^{n + 1 -i} S_{u_{c_{ij}}}\Big],
\ea
while the readout noise PSD remains the same as Eq.\;(\ref{Sr}). There is an enhancement of $2^{2n-2}$ compared to the chain detector in Eq.\;(\ref{sigp}) with a single sensor. On the other hand, each mode of the intrinsic noise is incoherent and the PSD at each layer $i$ is counted by a factor of the number of the same type of the modes in that layer, $2^{n+1-i}$, compared to Eq.\;(\ref{Sint}).

Similarly, one optimizes and fits the $\widetilde{\textrm{SNR}}$ numerically at zero temperature
\ba
n=2&:& \gamma_r=1.15 (g^{4}\gamma)^{1/5}, \widetilde{\textrm{SNR}}=1.49 (g/\gamma)^{2/5};\nn\\
n=3&:& \gamma_r=1.21 (g^{6}\gamma)^{1/7}, \widetilde{\textrm{SNR}}=3.07 (g/\gamma)^{3/7},\nn\\\label{noptBT23}
\ea
which are shown in the lower panel of Fig.\;\ref{RCoptS}.
As expected, there is an $2^{n-1}$ enhancement due to the multi probing sensors:
\be \textrm{SNR} \simeq 2^{n-1} \left(\frac{g}{\gamma}\right)^{\frac{n}{2n+1}} \frac{\rho_{\textrm{DM}} \alpha^2}{m_\Phi^{3}} \sqrt{\frac{Q_\Phi t_e}{\gamma}}.\label{BTSNR}\ee

\begin{figure}[h]
  	\centering
  	\includegraphics[width=0.8\columnwidth]{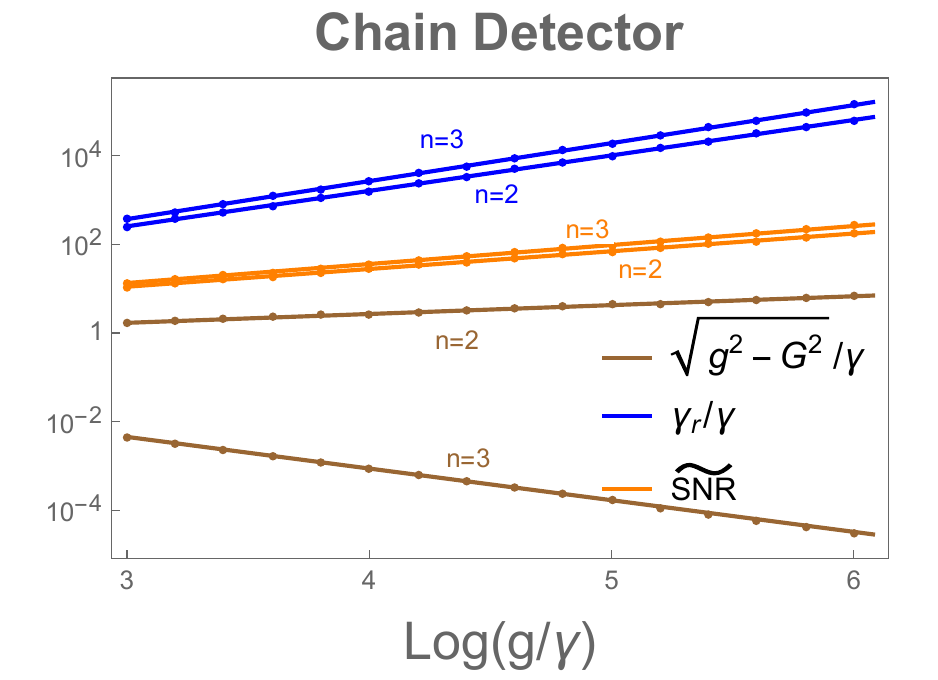}
  	\includegraphics[width=0.8\columnwidth]{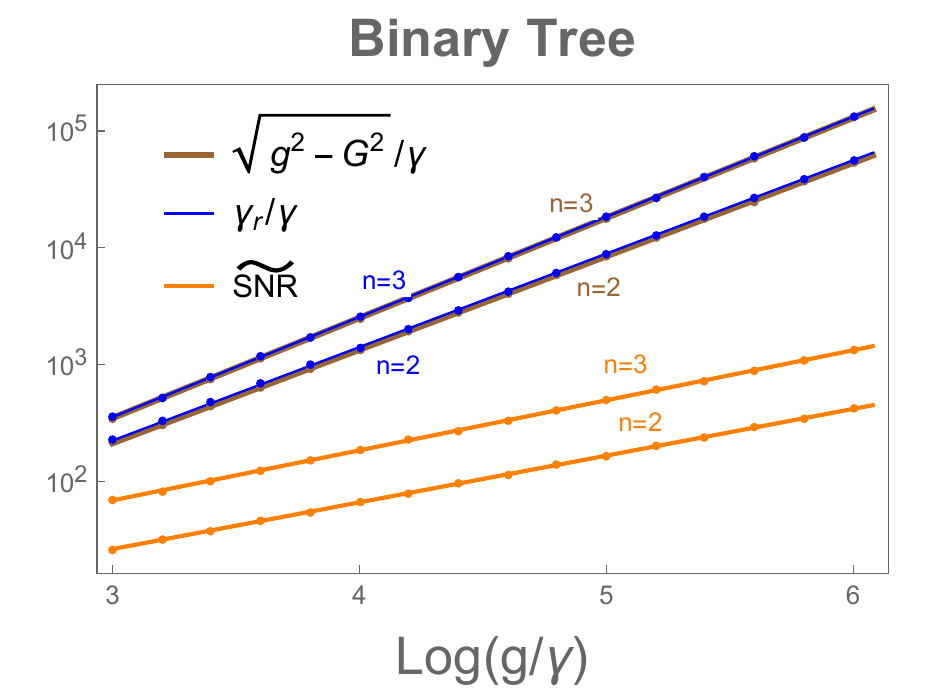}
  	\caption{
  		\footnotesize
  		{Optimization conditions and $\widetilde{\textrm{SNR}}$ with both the $\gamma_r$ and $g^2-G^2$ being free parameters for both chain detector and binary tree haloscope.}}
  	\label{Gopt}
  \end{figure}

We further optimize $\widetilde{\textrm{SNR}}$ with both $\gamma_r$ and $g^2 - G^2$ taken as free parameters. The results are shown in Fig.\;\ref{Gopt}, and the corresponding noise PSD is in Fig.\;\ref{nPToptPSD}. In the binary tree case, the optimized condition is $\sqrt{g^2-G^2} \simeq \gamma_r \simeq (g^{2n} \gamma)^{1/ (2n+1)}$. Thus a small deviation from the optimized value of $g^2-G^2$ can lead to negligible impact on the SNR and is indeed more robust than the chain detector when $g^2-G^2$ needs to be highly fine tuned.

In the presence of a large thermal occupation number $n_{\textrm{occ}} \equiv 1/2 + 1/(e^{\omega/T} -1)$, the thermal noise is dominant for each intrinsic noise PSD $S_\textrm{u} = n_{\textrm{occ}} \simeq T/\omega$. For SNR, the finite-temperature effect is equivalent to replacing $g^{2n}  \rightarrow g^{2n} n_{\textrm{occ}}$, with the optimized condition $\gamma_r \simeq (g^{2n} \gamma n_{\textrm{occ}})^{1/(2n+1)}$, and dividing the whole $\textrm{SNR}$ by $n_{\textrm{occ}}$. For the binary tree, Eq.\;(\ref{BTSNR}) becomes
\be \textrm{SNR} (T \gg \omega) \simeq 2^{n-1} \left(\frac{g}{\gamma n_{\textrm{occ}}}\right)^{\frac{n}{2n+1}} \frac{\rho_{\textrm{DM}} \alpha^2}{m_\Phi^{3}} \sqrt{\frac{Q_\Phi t_e}{\gamma n_{\textrm{occ}}}}.\label{BTSNRT}\ee
Notice that the thermal noise and the phase fluctuation noise for SRF both contain a frequency-dependent spectrum. Since our sensitivity bandwidth is largely broadened, one should consider this effect. The part with the frequency lower than the the center resonant frequency contributes more while the higher-frequency part contributes less. Thus evaluating the spectrum at the center resonant frequency serves as a viable approximation.

In the case of a microwave cavity with a dc magnetic field, the thermal noise is negligible with $T=10$ mK and
\be \textrm{SNR}_{\textrm{cavity}} = 2^{n-1} \left(\frac{g}{\gamma}\right)^{\frac{n}{2n+1}}  \frac{\rho_{\textrm{DM}} g_{\Phi\gamma}^2 \eta^2 B_0^2 V}{m_\Phi^{2}} \sqrt{\frac{Q_\Phi Q_{\textrm{int}} t_e}{m_\Phi}}.\ee
For the $LC$ circuit and SRF case, the SNR in the thermal noise limit are given by
\ba
\textrm{SNR}_{\textrm{LC}}& =& 2^{n-1} \left(\frac{g\, m_{\Phi}}{\gamma\, T}\right)^{\frac{n}{2n+1}} V^{\frac{5}{3}}g^2_{\Phi\gamma}B^2_{0}\rho_{\textrm{DM}}\sqrt{\frac{Q_{\Phi}Q_{\textrm{int}} t_{e}}{T}},\nn\\
\textrm{SNR}_{\textrm{SRF}}& =& 2^{n-1} \left(\frac{g\, \omega_{\textrm{rf}}}{\gamma\, T}\right)^{\frac{n}{2n+1}}  \frac{\rho_{\textrm{DM}} g_{\Phi\gamma}^2 \eta^2 B_0^2 V}{m_\Phi\ \omega_{\textrm{rf}}} \sqrt{\frac{Q_{a} Q_{\mathrm{int}} t_{e}}{T}},\nn\\\label{SNRSRF}
\ea
respectively.

{In the low frequency, the scan bandwidth $\Delta\omega_{\rm sc} \simeq (g^{2n} \gamma n_{\textrm{occ}})^{1/(2n+1)}$ becomes larger than the axion mass $m_\Phi$ for SRF with $g = \omega_{\textrm{rf}}$. In such cases, the enhancement to the scan rate from the increasing bandwidth will be limited, otherwise the integration time defined as $t_{\textrm{int}} = t_e \Delta \omega_{\rm sc}/m_\Phi$ will be much larger than the $e$-fold time $t_e$.
Thus we require the scan bandwidth $\Delta\omega_{\rm sc}$ to saturate $m_\Phi$ so that the integration time $t_{\textrm{int}} = t_e$. Thus the scaling with $m_\Phi$ changes below $10$ MHz for Eq.\;(\ref{SNRSRF}) and Fig.\;\ref{PR}.}


\begin{thebibliography}{10}

\bibitem{Peccei:1977hh}
R.D. Peccei and Helen~R. Quinn.
\newblock {CP Conservation in the Presence of Instantons}.
\newblock {\em Phys. Rev. Lett.}, 38:1440--1443, 1977.

\bibitem{Arvanitaki:2009fg}
Asimina Arvanitaki, Savas Dimopoulos, Sergei Dubovsky, Nemanja Kaloper, and
  John March-Russell.
\newblock {String Axiverse}.
\newblock {\em Phys. Rev. D}, 81:123530, 2010.

\bibitem{Preskill:1982cy}
John Preskill, Mark~B. Wise, and Frank Wilczek.
\newblock {Cosmology of the Invisible Axion}.
\newblock {\em Phys. Lett. B}, 120:127--132, 1983.

\bibitem{Abbott:1982af}
L.F. Abbott and P.~Sikivie.
\newblock {A Cosmological Bound on the Invisible Axion}.
\newblock {\em Phys. Lett. B}, 120:133--136, 1983.

\bibitem{Dine:1982ah}
Michael Dine and Willy Fischler.
\newblock {The Not So Harmless Axion}.
\newblock {\em Phys. Lett. B}, 120:137--141, 1983.

\bibitem{Sikivie:1983ip}
P.~Sikivie.
\newblock {Experimental Tests of the Invisible Axion}.
\newblock {\em Phys. Rev. Lett.}, 51:1415--1417, 1983.
\newblock [Erratum: Phys.Rev.Lett. 52, 695 (1984)].

\bibitem{Sikivie:1985yu}
Pierre Sikivie.
\newblock {Detection Rates for 'Invisible' Axion Searches}.
\newblock {\em Phys. Rev. D}, 32:2988, 1985.
\newblock [Erratum: Phys.Rev.D 36, 974 (1987)].

\bibitem{Sikivie:2013laa}
P.~Sikivie, N.~Sullivan, and D.B. Tanner.
\newblock {Proposal for Axion Dark Matter Detection Using an LC Circuit}.
\newblock {\em Phys. Rev. Lett.}, 112(13):131301, 2014.

\bibitem{Carroll:1989vb}
Sean~M. Carroll, George~B. Field, and Roman Jackiw.
\newblock {Limits on a Lorentz and Parity Violating Modification of
  Electrodynamics}.
\newblock {\em Phys. Rev.}, D41:1231, 1990.

\bibitem{Harari:1992ea}
Diego Harari and Pierre Sikivie.
\newblock {Effects of a Nambu-Goldstone boson on the polarization of radio
  galaxies and the cosmic microwave background}.
\newblock {\em Phys. Lett. B}, 289:67--72, 1992.

\bibitem{Chen:2019fsq}
Yifan Chen, Jing Shu, Xiao Xue, Qiang Yuan, and Yue Zhao.
\newblock {Probing Axions with Event Horizon Telescope Polarimetric
  Measurements}.
\newblock {\em Phys. Rev. Lett.}, 124(6):061102, 2020.

\bibitem{Graham:2013gfa}
Peter~W. Graham and Surjeet Rajendran.
\newblock {New Observables for Direct Detection of Axion Dark Matter}.
\newblock {\em Phys. Rev.}, D88:035023, 2013.

\bibitem{Budker:2013hfa}
Dmitry Budker, Peter~W. Graham, Micah Ledbetter, Surjeet Rajendran, and Alex
  Sushkov.
\newblock {Proposal for a Cosmic Axion Spin Precession Experiment (CASPEr)}.
\newblock {\em Phys. Rev.}, X4(2):021030, 2014.

\bibitem{Jiang:2021dby}
Min Jiang, Haowen Su, Antoine Garcon, Xinhua Peng, and Dmitry Budker.
\newblock {Search for axion-like dark matter with spin-based amplifiers}.
\newblock {\em Nature Phys.}, 17(12):1402--1407, 2021.

\bibitem{Du:2018uak}
N.~Du et~al.
\newblock {A Search for Invisible Axion Dark Matter with the Axion Dark Matter
  Experiment}.
\newblock {\em Phys. Rev. Lett.}, 120(15):151301, 2018.

\bibitem{Payez:2014xsa}
Alexandre Payez, Carmelo Evoli, Tobias Fischer, Maurizio Giannotti, Alessandro
  Mirizzi, and Andreas Ringwald.
\newblock {Revisiting the SN1987A gamma-ray limit on ultralight axion-like
  particles}.
\newblock {\em JCAP}, 02:006, 2015.

\bibitem{Anastassopoulos:2017ftl}
V.~Anastassopoulos et~al.
\newblock {New CAST Limit on the Axion-Photon Interaction}.
\newblock {\em Nature Phys.}, 13:584--590, 2017.

\bibitem{McAllister:2017lkb}
Ben~T. McAllister, Graeme Flower, Eugene~N. Ivanov, Maxim Goryachev, Jeremy
  Bourhill, and Michael~E. Tobar.
\newblock {The ORGAN Experiment: An axion haloscope above 15 GHz}.
\newblock {\em Phys. Dark Univ.}, 18:67--72, 2017.

\bibitem{Brubaker:2018ebj}
Benjamin~M. Brubaker.
\newblock {\em {First results from the HAYSTAC axion search}}.
\newblock PhD thesis, Yale U., 2017.

\bibitem{Lee:2020cfj}
S.~Lee, S.~Ahn, J.~Choi, B.~R. Ko, and Y.~K. Semertzidis.
\newblock {Axion Dark Matter Search around 6.7 $\mu$eV}.
\newblock {\em Phys. Rev. Lett.}, 124(10):101802, 2020.

\bibitem{Chaudhuri:2014dla}
Saptarshi Chaudhuri, Peter~W. Graham, Kent Irwin, Jeremy Mardon, Surjeet
  Rajendran, and Yue Zhao.
\newblock {Radio for hidden-photon dark matter detection}.
\newblock {\em Phys. Rev. D}, 92(7):075012, 2015.

\bibitem{Ouellet:2018beu}
Jonathan~L. Ouellet et~al.
\newblock {First Results from ABRACADABRA-10 cm: A Search for Sub-$\mu$eV Axion
  Dark Matter}.
\newblock {\em Phys. Rev. Lett.}, 122(12):121802, 2019.

\bibitem{Berlin:2019ahk}
Asher Berlin, Raffaele~Tito D'Agnolo, Sebastian~A.R. Ellis, Christopher
  Nantista, Jeffrey Neilson, Philip Schuster, Sami Tantawi, Natalia Toro, and
  Kevin Zhou.
\newblock {Axion Dark Matter Detection by Superconducting Resonant Frequency
  Conversion}.
\newblock {\em JHEP}, 07(07):088, 2020.

\bibitem{Lasenby:2019prg}
Robert Lasenby.
\newblock {Microwave cavity searches for low-frequency axion dark matter}.
\newblock {\em Phys. Rev. D}, 102(1):015008, 2020.

\bibitem{Lasenby:2019hfz}
Robert Lasenby.
\newblock {Parametrics of Electromagnetic Searches for Axion Dark Matter}.
\newblock {\em Phys. Rev. D}, 103(7):075007, 2021.

\bibitem{Malnou:2018dxn}
M.~Malnou, D.~A. Palken, B.~M. Brubaker, Leila~R. Vale, Gene~C. Hilton, and
  K.~W. Lehnert.
\newblock {Squeezed vacuum used to accelerate the search for a weak classical
  signal}.
\newblock {\em Phys. Rev. X}, 9(2):021023, 2019.
\newblock [Erratum: Phys.Rev.X 10, 039902 (2020)].

\bibitem{Chaudhuri:2018rqn}
Saptarshi Chaudhuri, Kent Irwin, Peter~W. Graham, and Jeremy Mardon.
\newblock {Optimal Impedance Matching and Quantum Limits of Electromagnetic
  Axion and Hidden-Photon Dark Matter Searches}.
\newblock{\it arXiv preprint arXiv:1803.01627 }

\bibitem{Chaudhuri:2019ntz}
Saptarshi Chaudhuri, Kent~D. Irwin, Peter~W. Graham, and Jeremy Mardon.
\newblock {Optimal Electromagnetic Searches for Axion and Hidden-Photon Dark
  Matter}.
\newblock{\it arXiv preprint arXiv:1904.05806 }

\bibitem{Krauss:1985ub}
Lawrence Krauss, John Moody, Frank Wilczek, and Donald~E. Morris.
\newblock {Calculations for Cosmic Axion Detection}.
\newblock {\em Phys. Rev. Lett.}, 55:1797, 1985.

\bibitem{Zheng:2016qjv}
Huaixiu Zheng, Matti Silveri, R.T. Brierley, S.M. Girvin, and K.W. Lehnert.
\newblock {Accelerating dark-matter axion searches with quantum measurement
  technology}.
\newblock{\it arXiv preprint arXiv:1607.02529 }

\bibitem{Berlin:2020vrk}
Asher Berlin, Raffaele~Tito D'Agnolo, Sebastian A.~R. Ellis, and Kevin Zhou.
\newblock {Heterodyne broadband detection of axion dark matter}.
\newblock {\em Phys. Rev. D}, 104(11):L111701, 2021.

\bibitem{Li:2020cwh}
Xiang Li, Maxim Goryachev, Yiqiu Ma, Michael~E. Tobar, Chunnong Zhao, Rana~X.
  Adhikari, and Yanbei Chen.
\newblock {Broadband sensitivity improvement via coherent quantum feedback with
  PT symmetry}.
\newblock{\it arXiv preprint arXiv:2012.00836 }

\bibitem{Miao:2015pna}
Haixing Miao, Yiqiu Ma, Chunnong Zhao, and Yanbei Chen.
\newblock {Enhancing the bandwidth of gravitational-wave detectors with
  unstable optomechanical filters}.
\newblock {\em Phys. Rev. Lett.}, 115(21):211104, 2015.

\bibitem{Derevianko:2016vpm}
Andrei Derevianko.
\newblock {Detecting dark-matter waves with a network of precision-measurement
  tools}.
\newblock {\em Phys. Rev. A}, 97(4):042506, 2018.

\bibitem{Foster:2020fln}
Joshua~W. Foster, Yonatan Kahn, Rachel Nguyen, Nicholas~L. Rodd, and
  Benjamin~R. Safdi.
\newblock {Dark Matter Interferometry}.
\newblock {\em Phys. Rev. D}, 103(7):076018, 2021.

\bibitem{Chen:2021bdr}
Yifan Chen, Min Jiang, Jing Shu, Xiao Xue, and Yanjie Zeng.
\newblock {Dissecting Axion and Dark Photon with A Network of Vector Sensors}.
\newblock{\it arXiv preprint arXiv:2111.06732 }

\bibitem{PCLC}
Johannes Russer and Peter Russer.
\newblock Circuit models in quantum electrodynamics.
\newblock In {\em 2011 XXXth URSI General Assembly and Scientific Symposium},
  pages 1--4, 2011.

\bibitem{Bergeal}
N.~Bergeal, R.~Vijay, V.~E. Manucharyan, I.~Siddiqi, R.~J. Schoelkopf, S.~M.
  Girvin, and M.~H. Devoret.
\newblock Analog information processing at the quantum limit with a josephson
  ring modulator.
\newblock {\em Nature Physics}, 6(4):296--302, 2010.

\bibitem{Wurtz:2021cnm}
K.~Wurtz, B.~M. Brubaker, Y.~Jiang, E.~P. Ruddy, D.~A. Palken, and K.~W.
  Lehnert.
\newblock {Cavity Entanglement and State Swapping to Accelerate the Search for
  Axion Dark Matter}.
\newblock {\em PRX Quantum}, 2(4):040350, 2021.

\bibitem{WBAR}
Hui Wang, M.~P. Blencowe, A.~D. Armour, and A.~J. Rimberg.
\newblock Quantum dynamics of a josephson junction driven cavity mode system in
  the presence of voltage bias noise.
\newblock {\em Physical Review B}, 96(10), Sep 2017.

\end{thebibliography}
\end{document}